\documentclass[letterpaper,compsoc,twoside]{IEEEtran}
\usepackage{fixltx2e} 
\usepackage{cmap} 
\usepackage{ifthen}
\usepackage[T1]{fontenc}
\usepackage[utf8]{inputenc}
\usepackage{amsmath}

\usepackage[font={small,it},labelfont=bf]{caption}
\usepackage{float}

\usepackage{nicefrac}
\usepackage{longtable,ltcaption,array}
\newlength{\DUtablewidth} 

\pdfoutput=1
\usepackage{scipy}
\makeatletter
\def\PY@reset{\let\PY@it=\relax \let\PY@bf=\relax%
    \let\PY@ul=\relax \let\PY@tc=\relax%
    \let\PY@bc=\relax \let\PY@ff=\relax}
\def\PY@tok#1{\csname PY@tok@#1\endcsname}
\def\PY@toks#1+{\ifx\relax#1\empty\else%
    \PY@tok{#1}\expandafter\PY@toks\fi}
\def\PY@do#1{\PY@bc{\PY@tc{\PY@ul{%
    \PY@it{\PY@bf{\PY@ff{#1}}}}}}}
\def\PY#1#2{\PY@reset\PY@toks#1+\relax+\PY@do{#2}}

\expandafter\def\csname PY@tok@gd\endcsname{\def\PY@tc##1{\textcolor[rgb]{0.63,0.00,0.00}{##1}}}
\expandafter\def\csname PY@tok@gu\endcsname{\let\PY@bf=\textbf\def\PY@tc##1{\textcolor[rgb]{0.50,0.00,0.50}{##1}}}
\expandafter\def\csname PY@tok@gt\endcsname{\def\PY@tc##1{\textcolor[rgb]{0.00,0.27,0.87}{##1}}}
\expandafter\def\csname PY@tok@gs\endcsname{\let\PY@bf=\textbf}
\expandafter\def\csname PY@tok@gr\endcsname{\def\PY@tc##1{\textcolor[rgb]{1.00,0.00,0.00}{##1}}}
\expandafter\def\csname PY@tok@cm\endcsname{\let\PY@it=\textit\def\PY@tc##1{\textcolor[rgb]{0.25,0.50,0.56}{##1}}}
\expandafter\def\csname PY@tok@vg\endcsname{\def\PY@tc##1{\textcolor[rgb]{0.73,0.38,0.84}{##1}}}
\expandafter\def\csname PY@tok@m\endcsname{\def\PY@tc##1{\textcolor[rgb]{0.13,0.50,0.31}{##1}}}
\expandafter\def\csname PY@tok@mh\endcsname{\def\PY@tc##1{\textcolor[rgb]{0.13,0.50,0.31}{##1}}}
\expandafter\def\csname PY@tok@cs\endcsname{\def\PY@tc##1{\textcolor[rgb]{0.25,0.50,0.56}{##1}}\def\PY@bc##1{\setlength{\fboxsep}{0pt}\colorbox[rgb]{1.00,0.94,0.94}{\strut ##1}}}
\expandafter\def\csname PY@tok@ge\endcsname{\let\PY@it=\textit}
\expandafter\def\csname PY@tok@vc\endcsname{\def\PY@tc##1{\textcolor[rgb]{0.73,0.38,0.84}{##1}}}
\expandafter\def\csname PY@tok@il\endcsname{\def\PY@tc##1{\textcolor[rgb]{0.13,0.50,0.31}{##1}}}
\expandafter\def\csname PY@tok@go\endcsname{\def\PY@tc##1{\textcolor[rgb]{0.20,0.20,0.20}{##1}}}
\expandafter\def\csname PY@tok@cp\endcsname{\def\PY@tc##1{\textcolor[rgb]{0.00,0.44,0.13}{##1}}}
\expandafter\def\csname PY@tok@gi\endcsname{\def\PY@tc##1{\textcolor[rgb]{0.00,0.63,0.00}{##1}}}
\expandafter\def\csname PY@tok@gh\endcsname{\let\PY@bf=\textbf\def\PY@tc##1{\textcolor[rgb]{0.00,0.00,0.50}{##1}}}
\expandafter\def\csname PY@tok@ni\endcsname{\let\PY@bf=\textbf\def\PY@tc##1{\textcolor[rgb]{0.84,0.33,0.22}{##1}}}
\expandafter\def\csname PY@tok@nl\endcsname{\let\PY@bf=\textbf\def\PY@tc##1{\textcolor[rgb]{0.00,0.13,0.44}{##1}}}
\expandafter\def\csname PY@tok@nn\endcsname{\let\PY@bf=\textbf\def\PY@tc##1{\textcolor[rgb]{0.05,0.52,0.71}{##1}}}
\expandafter\def\csname PY@tok@no\endcsname{\def\PY@tc##1{\textcolor[rgb]{0.38,0.68,0.84}{##1}}}
\expandafter\def\csname PY@tok@na\endcsname{\def\PY@tc##1{\textcolor[rgb]{0.25,0.44,0.63}{##1}}}
\expandafter\def\csname PY@tok@nb\endcsname{\def\PY@tc##1{\textcolor[rgb]{0.00,0.44,0.13}{##1}}}
\expandafter\def\csname PY@tok@nc\endcsname{\let\PY@bf=\textbf\def\PY@tc##1{\textcolor[rgb]{0.05,0.52,0.71}{##1}}}
\expandafter\def\csname PY@tok@nd\endcsname{\let\PY@bf=\textbf\def\PY@tc##1{\textcolor[rgb]{0.33,0.33,0.33}{##1}}}
\expandafter\def\csname PY@tok@ne\endcsname{\def\PY@tc##1{\textcolor[rgb]{0.00,0.44,0.13}{##1}}}
\expandafter\def\csname PY@tok@nf\endcsname{\def\PY@tc##1{\textcolor[rgb]{0.02,0.16,0.49}{##1}}}
\expandafter\def\csname PY@tok@si\endcsname{\let\PY@it=\textit\def\PY@tc##1{\textcolor[rgb]{0.44,0.63,0.82}{##1}}}
\expandafter\def\csname PY@tok@s2\endcsname{\def\PY@tc##1{\textcolor[rgb]{0.25,0.44,0.63}{##1}}}
\expandafter\def\csname PY@tok@vi\endcsname{\def\PY@tc##1{\textcolor[rgb]{0.73,0.38,0.84}{##1}}}
\expandafter\def\csname PY@tok@nt\endcsname{\let\PY@bf=\textbf\def\PY@tc##1{\textcolor[rgb]{0.02,0.16,0.45}{##1}}}
\expandafter\def\csname PY@tok@nv\endcsname{\def\PY@tc##1{\textcolor[rgb]{0.73,0.38,0.84}{##1}}}
\expandafter\def\csname PY@tok@s1\endcsname{\def\PY@tc##1{\textcolor[rgb]{0.25,0.44,0.63}{##1}}}
\expandafter\def\csname PY@tok@gp\endcsname{\let\PY@bf=\textbf\def\PY@tc##1{\textcolor[rgb]{0.78,0.36,0.04}{##1}}}
\expandafter\def\csname PY@tok@sh\endcsname{\def\PY@tc##1{\textcolor[rgb]{0.25,0.44,0.63}{##1}}}
\expandafter\def\csname PY@tok@ow\endcsname{\let\PY@bf=\textbf\def\PY@tc##1{\textcolor[rgb]{0.00,0.44,0.13}{##1}}}
\expandafter\def\csname PY@tok@sx\endcsname{\def\PY@tc##1{\textcolor[rgb]{0.78,0.36,0.04}{##1}}}
\expandafter\def\csname PY@tok@bp\endcsname{\def\PY@tc##1{\textcolor[rgb]{0.00,0.44,0.13}{##1}}}
\expandafter\def\csname PY@tok@c1\endcsname{\let\PY@it=\textit\def\PY@tc##1{\textcolor[rgb]{0.25,0.50,0.56}{##1}}}
\expandafter\def\csname PY@tok@kc\endcsname{\let\PY@bf=\textbf\def\PY@tc##1{\textcolor[rgb]{0.00,0.44,0.13}{##1}}}
\expandafter\def\csname PY@tok@c\endcsname{\let\PY@it=\textit\def\PY@tc##1{\textcolor[rgb]{0.25,0.50,0.56}{##1}}}
\expandafter\def\csname PY@tok@mf\endcsname{\def\PY@tc##1{\textcolor[rgb]{0.13,0.50,0.31}{##1}}}
\expandafter\def\csname PY@tok@err\endcsname{\def\PY@bc##1{\setlength{\fboxsep}{0pt}\fcolorbox[rgb]{1.00,0.00,0.00}{1,1,1}{\strut ##1}}}
\expandafter\def\csname PY@tok@kd\endcsname{\let\PY@bf=\textbf\def\PY@tc##1{\textcolor[rgb]{0.00,0.44,0.13}{##1}}}
\expandafter\def\csname PY@tok@ss\endcsname{\def\PY@tc##1{\textcolor[rgb]{0.32,0.47,0.09}{##1}}}
\expandafter\def\csname PY@tok@sr\endcsname{\def\PY@tc##1{\textcolor[rgb]{0.14,0.33,0.53}{##1}}}
\expandafter\def\csname PY@tok@mo\endcsname{\def\PY@tc##1{\textcolor[rgb]{0.13,0.50,0.31}{##1}}}
\expandafter\def\csname PY@tok@mi\endcsname{\def\PY@tc##1{\textcolor[rgb]{0.13,0.50,0.31}{##1}}}
\expandafter\def\csname PY@tok@kn\endcsname{\let\PY@bf=\textbf\def\PY@tc##1{\textcolor[rgb]{0.00,0.44,0.13}{##1}}}
\expandafter\def\csname PY@tok@o\endcsname{\def\PY@tc##1{\textcolor[rgb]{0.40,0.40,0.40}{##1}}}
\expandafter\def\csname PY@tok@kr\endcsname{\let\PY@bf=\textbf\def\PY@tc##1{\textcolor[rgb]{0.00,0.44,0.13}{##1}}}
\expandafter\def\csname PY@tok@s\endcsname{\def\PY@tc##1{\textcolor[rgb]{0.25,0.44,0.63}{##1}}}
\expandafter\def\csname PY@tok@kp\endcsname{\def\PY@tc##1{\textcolor[rgb]{0.00,0.44,0.13}{##1}}}
\expandafter\def\csname PY@tok@w\endcsname{\def\PY@tc##1{\textcolor[rgb]{0.73,0.73,0.73}{##1}}}
\expandafter\def\csname PY@tok@kt\endcsname{\def\PY@tc##1{\textcolor[rgb]{0.56,0.13,0.00}{##1}}}
\expandafter\def\csname PY@tok@sc\endcsname{\def\PY@tc##1{\textcolor[rgb]{0.25,0.44,0.63}{##1}}}
\expandafter\def\csname PY@tok@sb\endcsname{\def\PY@tc##1{\textcolor[rgb]{0.25,0.44,0.63}{##1}}}
\expandafter\def\csname PY@tok@k\endcsname{\let\PY@bf=\textbf\def\PY@tc##1{\textcolor[rgb]{0.00,0.44,0.13}{##1}}}
\expandafter\def\csname PY@tok@se\endcsname{\let\PY@bf=\textbf\def\PY@tc##1{\textcolor[rgb]{0.25,0.44,0.63}{##1}}}
\expandafter\def\csname PY@tok@sd\endcsname{\let\PY@it=\textit\def\PY@tc##1{\textcolor[rgb]{0.25,0.44,0.63}{##1}}}

\def\PYZus{\char`\_}
\def\PYZob{\char`\{}
\def\PYZcb{\char`\}}

\def\PYZlt{\char`\<}
\def\PYZgt{\char`\>}
\def\PYZsh{\char`\#}

\def\PYZsq{\char`\'}

\makeatother

\providecommand*{\DUfootnotemark}[3]{%
  \raisebox{1em}{\hypertarget{#1}{}}%
  \hyperlink{#2}{\textsuperscript{#3}}%
}
\providecommand{\DUfootnotetext}[4]{%
  \begingroup%
  \renewcommand{\thefootnote}{%
    \protect\raisebox{1em}{\protect\hypertarget{#1}{}}%
    \protect\hyperlink{#2}{#3}}%
  \footnotetext{#4}%
  \endgroup%
}

\providecommand*{\DUrole}[2]{%
  \ifcsname DUrole#1\endcsname%
    \csname DUrole#1\endcsname{#2}%
  \else
    \ifcsname docutilsrole#1\endcsname%
      \csname docutilsrole#1\endcsname{#2}%
    \else%
      #2%
    \fi%
  \fi%
}

\ifthenelse{\isundefined{\hypersetup}}{
  \usepackage[colorlinks=true,linkcolor=blue,urlcolor=blue]{hyperref}
  \urlstyle{same} 
}{}

\begin{document}
\newcounter{footnotecounter}\title{SClib, a hack for straightforward embedded C functions in Python}\author{Esteban Fuentes$^{\setcounter{footnotecounter}{1}\fnsymbol{footnotecounter}\setcounter{footnotecounter}{2}\fnsymbol{footnotecounter}}$%
          \setcounter{footnotecounter}{1}\thanks{\fnsymbol{footnotecounter} %
          Corresponding author: \protect\href{mailto:esteban.fuentes@tum.de}{esteban.fuentes@tum.de}}\setcounter{footnotecounter}{2}\thanks{\fnsymbol{footnotecounter} Electric Drives and Power Electronics group at the Technical University of Munich}, Hector E. Martinez$^{\setcounter{footnotecounter}{3}\fnsymbol{footnotecounter}}$\setcounter{footnotecounter}{3}\thanks{\fnsymbol{footnotecounter} T30f group at the Physics Department of the Technical University of Munich}\thanks{%

          \noindent%
          Copyright\,\copyright\,2014 Esteban Fuentes et al. This is an open-access article distributed under the terms of the Creative Commons Attribution License, which permits unrestricted use, distribution, and reproduction in any medium, provided the original author and source are credited. http://creativecommons.org/licenses/by/3.0/%
        }}\maketitle
          \renewcommand{\leftmark}{PROC. OF THE 7th EUR. CONF. ON PYTHON IN SCIENCE (EUROSCIPY 2014)}
          \renewcommand{\rightmark}{SCLIB, A HACK FOR STRAIGHTFORWARD EMBEDDED C FUNCTIONS IN PYTHON}

\setcounter{page}{47}
\newcommand*{\docutilsroleref}{\ref}
\newcommand*{\docutilsrolelabel}{\label}
\AtEndDocument{\cleardoublepage}

\begin{abstract}We present SClib, a simple hack that allows easy and straightforward
evaluation of C functions within Python code, boosting flexibility for
better trade-off between computation power and feature availability, such as
visualization and existing computation routines in SciPy.

We also present two cases were SClib has been used.

In the first set of applications we use SClib to write a port to Python of a
Schrödinger equation solver that has been extensively used the literature,
the resulting script presents a speed-up of about 150x with respect to the original one.
A review of the situations where the speeded-up script has been used is presented.
We also describe the solution to the
related problem of solving a set of coupled Schrödinger-like equations
where SClib is used to implement the speed-critical parts of the code. We
argue that when using SClib within IPython we can use NumPy and Matplotlib
for the manipulation and visualization of the solutions in an interactive
environment with no performance compromise.

The second case is an engineering application. We use SClib to evaluate the
control and system derivatives in a feedback control loop for electrical
motors.  With this and the integration routines available in SciPy, we can
run simulations of the control loop a la Simulink. The use of C code not
only boosts the speed of the simulations, but also enables to test the
exact same code that we use in the test rig to get experimental results.
Again, integration with IPython gives us the flexibility to analyze and
visualize the data.\end{abstract}

\section{Introduction%
  \label{introduction}%
}

Embedding code written in oder languages is a common theme in the Python
context, the main motivation being speed boosting.
Several alternatives exist to achieve this, such as
Cython \cite{Cython}, CFFI \cite{CFFI}, SWIG \cite{SWIG}, weave \cite{weave}, among others.
We present yet another alternative, which may be quite close to CFFI than to
the others.
The motivation to write SClib grew out of the urge to integrate C code, which
was already written, into the Python environment, minimizing the intervention of
the code.
Part of the resulting work is briefly introduced later, in the engineering
application section.

Nevertheless, embedding compiled code in Python will naturally have an impact in
performance, for instance, when the compiled code takes care of computer
intensive numerics.  The first application we introduce (in particle physics),
leverages SClib in this sense, outsourcing the numerics to the compiled code and
using the Python environment for visualization.

\section{SClib%
  \label{sclib}%
}

At the core of SClib\DUfootnotemark{id5}{id7}{1} is ctypes \cite{Hell}, which actually does the whole
work: it maps Python data to C compatible data and provides a way to call
functions in DLLs or shared libraries.  SClib acts as glue: it puts things
together for the user, to provide him with an easy to use interface.%
\DUfootnotetext{id7}{id5}{1}{
The code for SClib and example use are available at <\url{https://github.com/drestebon/SClib}>}

The requirements for SClib are very simple: call a function on an array of
numbers of arbitrary type and size and return the output of the function, again
of arbitrary type and size.

The resulting interface is also very simple: A library is initialized in the
Python side with the path to the DLL (or shared library) and a list with the
names of the functions to be called:\begin{Verbatim}[commandchars=\\\{\},fontsize=\footnotesize]
\PY{n}{In} \PY{p}{[}\PY{l+m+mi}{1}\PY{p}{]}\PY{p}{:} \PY{k+kn}{import} \PY{n+nn}{SClib} \PY{k+kn}{as} \PY{n+nn}{sc}
\PY{n}{In} \PY{p}{[}\PY{l+m+mi}{2}\PY{p}{]}\PY{p}{:} \PY{n}{lib} \PY{o}{=} \PY{n}{sc}\PY{o}{.}\PY{n}{Clib}\PY{p}{(}\PY{l+s}{\PYZsq{}}\PY{l+s}{test.so}\PY{l+s}{\PYZsq{}}\PY{p}{,} \PY{p}{[}\PY{l+s}{\PYZsq{}}\PY{l+s}{fun}\PY{l+s}{\PYZsq{}}\PY{p}{]}\PY{p}{)}
\end{Verbatim}
The functions are then available as members of the library and can be called
with the appropriate number of arguments, which are one dimensional arrays of
numbers.  The function returns a list containing the output arrays of the
function:\begin{Verbatim}[commandchars=\\\{\},fontsize=\footnotesize]
\PY{n}{In} \PY{p}{[}\PY{l+m+mi}{3}\PY{p}{]}\PY{p}{:} \PY{n}{out}\PY{p}{,} \PY{o}{=} \PY{n}{lib}\PY{o}{.}\PY{n}{fun}\PY{p}{(}\PY{p}{[}\PY{l+m+mi}{0}\PY{p}{]}\PY{p}{)}
\end{Verbatim}
In the C counterpart, the function declaration must be accompanied with
specifications of the inputs and outputs lengths and types. This is
accomplished with the helper macros defined in sclib.h:\begin{Verbatim}[commandchars=\\\{\},fontsize=\footnotesize]
\PY{c+cp}{\PYZsh{}}\PY{c+cp}{include \PYZlt{}sclib.h\PYZgt{}}
\PY{n}{SCL\PYZus{}OL}\PY{p}{(}\PY{n}{fun}\PY{p}{,} \PY{l+m+mi}{1}\PY{p}{,}   \PY{l+m+mi}{1}\PY{p}{)}\PY{p}{;}   \PY{c+cm}{/* outputs lengths */}
\PY{n}{SCL\PYZus{}OT}\PY{p}{(}\PY{n}{fun}\PY{p}{,} \PY{l+m+mi}{1}\PY{p}{,} \PY{n}{INT}\PY{p}{)}\PY{p}{;}   \PY{c+cm}{/* outputs types */}
\PY{n}{SCL\PYZus{}IL}\PY{p}{(}\PY{n}{fun}\PY{p}{,} \PY{l+m+mi}{1}\PY{p}{,}   \PY{l+m+mi}{1}\PY{p}{)}\PY{p}{;}   \PY{c+cm}{/* inputs lengths */}
\PY{n}{SCL\PYZus{}IT}\PY{p}{(}\PY{n}{fun}\PY{p}{,} \PY{l+m+mi}{1}\PY{p}{,} \PY{n}{INT}\PY{p}{)}\PY{p}{;}   \PY{c+cm}{/* inputs types */}
\PY{k+kt}{void} \PY{n+nf}{fun}\PY{p}{(}\PY{k+kt}{int} \PY{o}{*} \PY{n}{out}\PY{p}{,} \PY{k+kt}{int} \PY{o}{*} \PY{n}{in}\PY{p}{)} \PY{p}{\PYZob{}}
    \PY{o}{*}\PY{n}{out} \PY{o}{=} \PY{l+m+mi}{42}\PY{p}{;}
\PY{p}{\PYZcb{}}
\end{Verbatim}
An arbitrary number of inputs or outputs can be specified, for example:\begin{Verbatim}[commandchars=\\\{\},fontsize=\footnotesize]
\PY{c+cp}{\PYZsh{}}\PY{c+cp}{include \PYZlt{}math.h\PYZgt{}}
\PY{c+cp}{\PYZsh{}}\PY{c+cp}{include \PYZlt{}sclib.h\PYZgt{}}
\PY{n}{SCL\PYZus{}OL}\PY{p}{(}\PY{n}{fun}\PY{p}{,} \PY{l+m+mi}{2}\PY{p}{,}   \PY{l+m+mi}{1}\PY{p}{,}     \PY{l+m+mi}{2}\PY{p}{)}\PY{p}{;}
\PY{n}{SCL\PYZus{}OT}\PY{p}{(}\PY{n}{fun}\PY{p}{,} \PY{l+m+mi}{2}\PY{p}{,} \PY{n}{INT}\PY{p}{,} \PY{n}{FLOAT}\PY{p}{)}\PY{p}{;}
\PY{n}{SCL\PYZus{}IL}\PY{p}{(}\PY{n}{fun}\PY{p}{,} \PY{l+m+mi}{2}\PY{p}{,}   \PY{l+m+mi}{1}\PY{p}{,}     \PY{l+m+mi}{2}\PY{p}{)}\PY{p}{;}
\PY{n}{SCL\PYZus{}IT}\PY{p}{(}\PY{n}{fun}\PY{p}{,} \PY{l+m+mi}{2}\PY{p}{,} \PY{n}{INT}\PY{p}{,} \PY{n}{FLOAT}\PY{p}{)}\PY{p}{;}
\PY{k+kt}{void} \PY{n+nf}{fun}\PY{p}{(}\PY{k+kt}{int} \PY{o}{*} \PY{n}{out0}\PY{p}{,} \PY{k+kt}{float} \PY{o}{*} \PY{n}{out1}\PY{p}{,}
         \PY{k+kt}{int} \PY{o}{*} \PY{n}{in0}\PY{p}{,} \PY{k+kt}{float} \PY{o}{*} \PY{n}{in1}\PY{p}{)} \PY{p}{\PYZob{}}
    \PY{o}{*}\PY{n}{out0} \PY{o}{=} \PY{l+m+mi}{42}\PY{o}{*}\PY{n}{in0}\PY{p}{[}\PY{l+m+mi}{0}\PY{p}{]}\PY{p}{;}
    \PY{n}{out1}\PY{p}{[}\PY{l+m+mi}{0}\PY{p}{]} \PY{o}{=} \PY{n}{in1}\PY{p}{[}\PY{l+m+mi}{0}\PY{p}{]}\PY{o}{*}\PY{n}{in1}\PY{p}{[}\PY{l+m+mi}{1}\PY{p}{]}\PY{p}{;}
    \PY{n}{out1}\PY{p}{[}\PY{l+m+mi}{1}\PY{p}{]} \PY{o}{=} \PY{n}{powf}\PY{p}{(}\PY{n}{in1}\PY{p}{[}\PY{l+m+mi}{0}\PY{p}{]}\PY{p}{,} \PY{n}{in1}\PY{p}{[}\PY{l+m+mi}{1}\PY{p}{]}\PY{p}{)}\PY{p}{;}
\PY{p}{\PYZcb{}}
\end{Verbatim}
In the function declaration, all the outputs must precede the inputs and must
be placed in the same order as in the SCL macros.

These specifications are processed during compilation time, but only the number
of inputs and outputs is static, the lengths of each component can be
overridden at run time:\begin{Verbatim}[commandchars=\\\{\},fontsize=\footnotesize]
\PY{n}{In} \PY{p}{[}\PY{l+m+mi}{4}\PY{p}{]}\PY{p}{:} \PY{n}{lib}\PY{o}{.}\PY{n}{INPUT\PYZus{}LEN}\PY{p}{[}\PY{l+s}{\PYZsq{}}\PY{l+s}{fun}\PY{l+s}{\PYZsq{}}\PY{p}{]} \PY{o}{=} \PY{p}{[}\PY{l+m+mi}{10}\PY{p}{,} \PY{l+m+mi}{1}\PY{p}{]}
\PY{n}{In} \PY{p}{[}\PY{l+m+mi}{5}\PY{p}{]}\PY{p}{:} \PY{n}{lib}\PY{o}{.}\PY{n}{retype}\PY{p}{(}\PY{p}{)}
\end{Verbatim}
In these use cases the length of the arguments should be given to the function
through an extra integer argument.

In the function body, both inputs and outputs should be treated as one
dimensional arrays.

\section{Application in Quarkonium Physics%
  \label{application-in-quarkonium-physics}%
}

\subsection{Motivation%
  \label{motivation}%
}

The Schrödinger equation is the fundamental equation for
describing non-relativistic quantum mechanical dynamics. For the applications
we will present in this section we will focus on the time-independent version
which, in natural units, is given by\begin{equation}
\label{schroe}
\left(-\frac{\nabla_{\mathbf r}^2}{2\mu}+V(\mathbf{r})\right)\psi(\mathbf{r}) = E\psi(\mathbf{r}).
\end{equation}It corresponds to an eigenvalue equation where the term inside the parenthesis
in l.h.s. is called the Hamiltonian operator, the value $E$, its
eigenvalue, is the measurable quantity (the energy) associated with it,
$\mu$ is the reduced mass of the system  (it correspond the mass of the particle in one-particle systems)
and the wavefunction, $\psi(\mathbf{r})$ is the entity
containing all the information about the system, since its modulus squared
correspond to the probability density of a given measurement, it has to be
normalized to unity. The term $V({\mathbf r})$ in the Hamiltonian is
called the potential.

Since its discovery, the Schrödinger equation has played an important role in
our understanding of nature and it is present in almost every aspect of modern
physics. In this section we will review some cases where SClib has been used to
implement solutions of the computing problems associated with eq.
(\DUrole{ref}{schroe}) that arise in the study of heavy quarkonia\DUfootnotemark{id8}{id9}{2}.%
\DUfootnotetext{id9}{id8}{2}{
For a comprehensive review of the status and perspectives of the
research in heavy quarkonia we refer the reader to chapter four of
\cite{Bra14}.}

Quarkonium is a bound-state composed by a quark and its corresponding
antiquark. By heavy we mean states composed by charm and bottom quarks,
called charmonium and bottomonium respectively. Due to its large mass, the top
quark decays before forming a bound state. In heavy quarkonium the relative
velocity between the quark and antiquark inside of the bound-system is believed
to be small enough for the system to be considered, at least in a first
approximation, non-relativistic, making it suitable for being described by eq.
(\DUrole{ref}{schroe}). Considering the equal mass case with a spherically symmetric
potential the angular part can be neglected (it correspond to the spherical
harmonics) and the relevant part of eq. (\DUrole{ref}{schroe}) reduces to the
one-dimensional equation given by\begin{equation}
\label{reduced}
\left[-\frac{1}{m}\frac{d^2}{dr^2}+\frac{l(l+1)}{mr^2}+V(r)\right]y_{n,l}(r)=E_{n,l}y_{n,l}(r),
\end{equation}where $r$ is the relative distance between the quark and the antiquark,
$l$ is the angular momentum quantum number, $m$ is the (anti)quark
mass (for equal mass $2\mu=m$), $y_{n,l}$ is called the reduced wavefunction and the eigenvalue
$E_{n,l}$ is interpreted as the binding energy of the bound-system, where
$n=0,1,2,\dots$ accounts for the number of nodes (radial excitations) of
the wavefunction. The mass of the quarkonium is then given by\begin{equation}
\label{lomass}
 M=2m+E_{n,l}.
\end{equation}The potential $V(r)$ describes the quark-antiquark interaction, it is a
function of $r$ and $\Lambda_{\rm QCD}$, the typical hadronic scale
($\sim 200\,{\rm MeV}$). For $r\Lambda_{\rm QCD} \ll 1$
(short-distance regime) the potential may be evaluated perturbatively, but for
$r\Lambda_{\rm QCD} \sim 1$ (long-distance regime) it cannot. To
overcome this issue, models based on non-relativistic reductions of
phenomenological observations have been used to describe heavy quarkonia, one
these being the so-called Cornell potential
\cite{Eich74}, \cite{Eich78}, \cite{Eich79}\begin{equation}
\label{cornell}
V(r) = \frac{a}{r}+kr,
\end{equation}where $a$ and $k$ are parameters which need to be fixed by
experimental (or lattice) data of some observable. This potential incorporates
two of the main observed characteristics of the quark-antiquark interaction: at
short distances it exhibits a Coulombic behavior and in the long-distance
regime the interaction is dominated by a confinement phase.

Since the beginning of the last decade, non-relativistic effective field
theories (EFT), in particular non-relativistic QCD (NRQCD) \cite{Cas85}, \cite{Bod94}
and potential NRQCD (pNRQCD) \cite{Bra99}, have become the state-of-the-art tools
for the study of heavy quarkonia (for review see \cite{Bra04}).  NRQCD is obtained
from QCD integrating out modes that scale like $m$, while pNRQCD is
obtained from NRQCD integrating out modes that scale like the quark momentum\DUfootnotemark{id18}{id19}{3}.%
\DUfootnotetext{id19}{id18}{3}{
These EFT exploit the hierarchy of energy scales present in the
bound-system. If the relative velocity of the (anti)quark,
$v$, is small, we have that $mv^2(\sim E)\ll mv(\sim p) \ll m$,
where $p$ is the momentum of the particles and $E$ its kinetic
energy. If one is interested in studying a phenomena that happens at the scale
$E$ (like the binding) it is more suitable to integrate out degrees of
freedoms with energies that scale like the other two higher scales, this is the
motivation behind pNRQCD. For a detailed analysis of the scales present in heavy
quarkonia we refer the reader to \cite{Bra04}.}

The physics of the modes that have been integrated out is encoded in Wilson
coefficients that must be calculated comparing at the same  scale the results
(observables, Green functions) of the EFT, with the ones of QCD (for NRQCD) or
NRQCD (for pNRQCD). A key feature of pNRQCD is that it allows the relativistic
corrections to the quark-antiquark potential to be organized as an expansion in
powers of $1/m$. Up to second order $V(r)$ can be written as\begin{equation}
\label{pnrqcdpot}
V(r)=V^{(0)}(r)+\frac{V^{(1/m)}(r)}{m}+\frac{V^{(1/m^2)}(r)}{m^2},
\end{equation}where $V^{(1/m)}$ and $V^{(1/m^2)}$ are derived from QCD through
the matching procedure with NRQCD. The details about $V^{(1/m)}$ and
$V^{(1/m^2)}$  and how they are obtained are beyond the scope of this
document, however, we can list some of their features:%
\begin{itemize}

\item 

They correspond to correlators that in the short-distance regime can
be computed in perturbation theory.
\item 

In the long-distance regime they can be computed in lattice QCD,
however only some of these correlators have been calculated.
\item 

Eq. (\DUrole{ref}{cornell}) correspond, at least qualitatively, to the leading
order $V^{(0)}$ in eq. (\DUrole{ref}{pnrqcdpot})
\end{itemize}

For the details about the derivation of the terms present in eq.
(\DUrole{ref}{pnrqcdpot}) we refer the reader to \cite{Bra00} and
\cite{Pin00}. It is important to recall that, although it can not be
evaluated analytically in the whole range of $r$, eq. (\DUrole{ref}{pnrqcdpot})
represents a model-independent expression for the quark-antiquark
potential, contrary to models like the one presented in eq. (\DUrole{ref}{cornell}).

Using perturbation theory to include the relativistic corrections to the potential, the expression for the
bound-state mass reads\begin{eqnarray}
\label{mass}
M&=&2m+E_{n,l}^{(0)}+\frac{\langle nl| V^{(1/m)}(r)|nl \rangle}{m}\\ \nonumber
 &+&\frac{\langle nl| V^{(1/m^2)}(r)|nl \rangle}{m^2}+\frac{1}{m^2}\sum_{m\neq n}^{\infty}\frac{| \langle nl|V^{(1/m)} | ml \rangle|^2}{E_{n,l}^{(0)}-E_{ml}^{(0)}},
\end{eqnarray}where $E_{il}^{(0)}$ comes from solving eq. (\DUrole{ref}{reduced}) with
$V(r)=V^{(0)}(r)$ and\begin{equation*}
\langle nl | f(r) | n'l' \rangle  \propto  \int_0^\infty dr\, y_{n,l}(r)f(r)y_{n'l'}(r),
\end{equation*}where the proportionality factor will depended on the corresponding quantum
numbers of the operators appearing in $V^{(1/m)}$ and $V^{(1/m^2)}$.\begin{figure}[htb]\noindent\makebox[\columnwidth][c]{\includegraphics[width=\columnwidth]{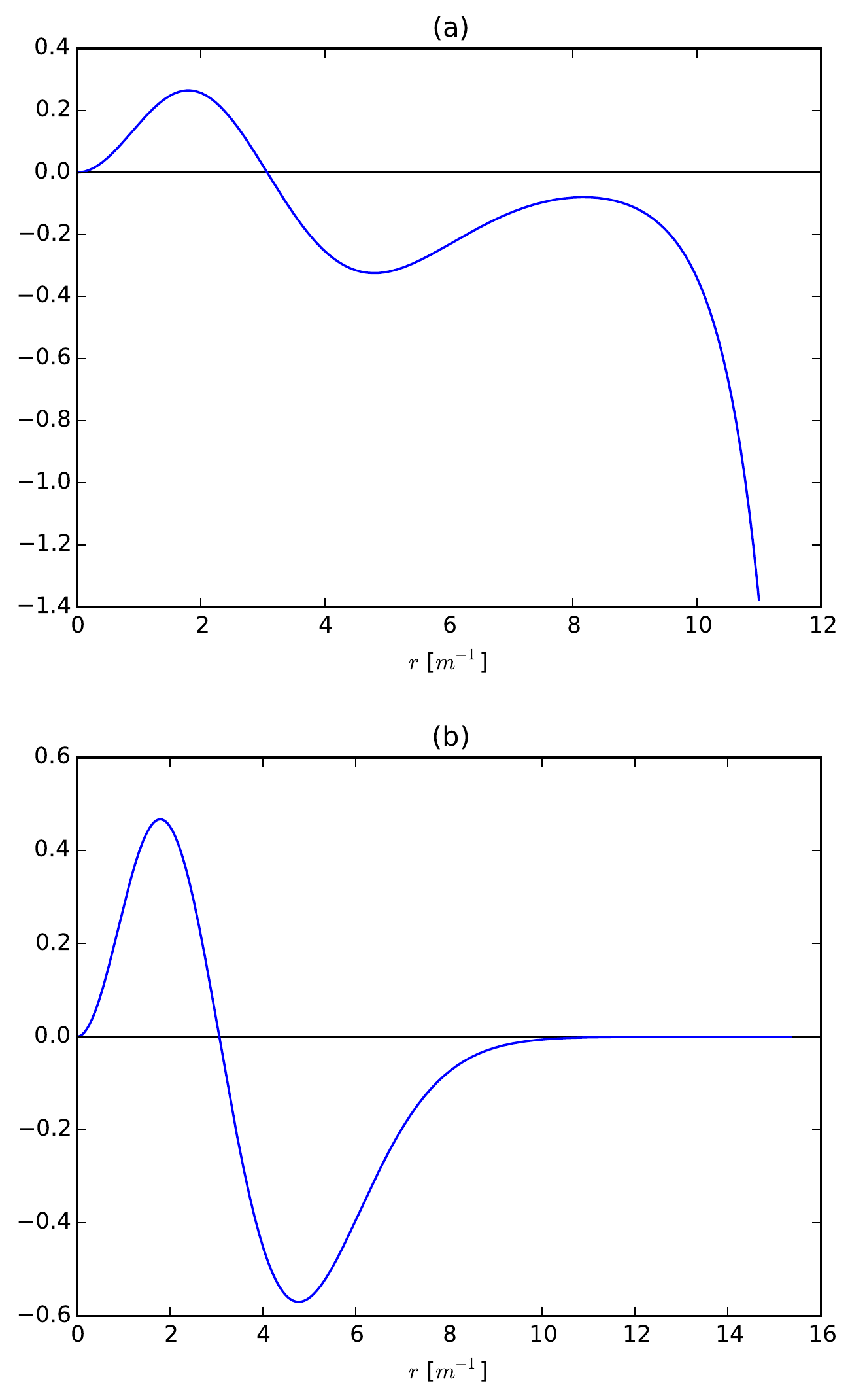}}
\caption{Reduced wavefunctions $y_{n,l}(r)$ for two steps in the search of the
eigenvalue $E_{n=1\,l=1}$. For each step in the process to find the
eigenvalue the nodes of the $y_{n,l}$ are counted, only when the
value of $E_{n,l}$ corresponds to an eigenvalue $y_{n,l}$ is not
divergent. In the plot (a) $E_{n,l} = 3.1\,m$ and $y_{n,l}$
fulfills the condition of having one node, however, the accuracy in the value
of $E_{n,l}$ is too low and the function diverges. In the plot (b)
$E_{n,l} = 3.10952\,m$ so $y_{n,l}\rightarrow 0$ for larger
values of $r$. We have used the Cornell potential eq. (\DUrole{ref}{cornell})
with parameters $m = 1$ $a = 0.1$, $k=0.5m^2$, all
dimensions defined in terms of the mass.}
\end{figure}

\subsection{Applications of SClib%
  \label{applications-of-sclib}%
}

The simplest computational problem related to eq. (\DUrole{ref}{reduced}) is to find
$E_{n,l}$ for a given $n$ and $l$. Methods to solve this
problem have been implemented since long ago (see for instance \cite{Fal85}), in a
nutshell, the standard method consist of applying two known constraints to the
reduced wavefunction $y_{n,l}$:%
\begin{itemize}

\item 

The number of nodes of $y_{n,l}(r)$ must be equal to $n$.
\item 

$y_{n,l}(r)$  has to be normalizable
\end{itemize}
\begin{equation}
\label{norm}
\int_0^\infty dr[y_{n,l}(r)]^2 = 1.
\end{equation}In general $y_{n,l}(r)$ will diverge except when $E_{n,l}$
corresponds to an eigenvalue. The procedure to find the eigenvalue consists in
to perform a scan of values of $E_{n,l}$ until $y_{n,l}(r)$  has
$n$ nodes and converges for a large enough value of $r$ (see Fig.
1). This implies that for each test value of $E_{n,l}$ eq.
(\DUrole{ref}{reduced}) must be (numerically) solved.  A popular\DUfootnotemark{id24}{id34}{4} Mathematica
\cite{Mat9} implementation of this method has been
available in \cite{Luc98}.  This script has the advantage that the user can profit
from the Mathematica built-in functions to plot, integrate or store the
resulting wavefunctions, however, it has a very poor performance.  With the
goal of mimicking some of the advantages of this script, but without compromising speed,
we ported the algorithm in \cite{Luc98} to Python. The resulting script, SChroe.py\DUfootnotemark{id28}{id35}{5}, uses SClib to implement
the speed-critical parts of the algorithm. In Schroe.py the wavefunctions are
stored as NumPy arrays \cite{NumPy} so when the script is run within IPython \cite{IPy}
together with SciPy \cite{SciPy}, NumPy and Matplotlib \cite{Mplot} the user can profit
of the same or more flexibility as with the Mathematica script plus a boosted
speed. In table 1 we compare the performance of SChroe.py against other
implementations of the same algorithm\DUfootnotemark{id33}{id36}{6}.%
\DUfootnotetext{id34}{id24}{4}{
The paper describing the script ranks fifth among the most cited papers
(91 citations) of the International Journal of Modern Physics C with the last
citation from  July 2014.}
\DUfootnotetext{id35}{id28}{5}{
Code available at <\url{https://github.com/heedmane/schroepy/}>}
\DUfootnotetext{id36}{id33}{6}{
Although the aim of this section is not to compare performance of Schrödinger equation solvers, but to present an application in which SClib can improve the speed of a known algorithm, we must mention that there are solvers that offer better performance than the current version of SChroe.py. For instance, the solver presented in \cite{dftatom} implements a more sophisticated integration method and allows refinements in the radial mesh. With these improvements the dftatom solver can reach a speed-up of at least two orders of magnitude compared to the current version of SChroe.py.}
\begin{table}
\setlength{\DUtablewidth}{0.8\linewidth}
\begin{longtable*}[c]{|p{0.133\DUtablewidth}|p{0.307\DUtablewidth}|p{0.249\DUtablewidth}|p{0.098\DUtablewidth}|p{0.156\DUtablewidth}|}
\hline

$n$ & 

$E_{n,l=1}\,\,[m]$ & 

schroe.nb \cite{Luc98} & 

Python & 

SChroe.py \\
\hline

0 & 

2.15789 & 

98.88 & 

25.46 & 

0.66 \\
\hline

1 & 

3.10952 & 

124.14 & 

30.95 & 

0.75 \\
\hline

2 & 

3.93850 & 

135.68 & 

35.32 & 

0.84 \\
\hline

20 & 

13.5995 & 

370.0 & 

88.04 & 

1.99 \\
\hline
\end{longtable*}
\caption{Time in seconds taken to compute the eigenvalues and reduced wavefunctions for the Cornell potential eq. (\DUrole{ref}{cornell}). The column Python correspond to the implementation of the algorithm in Python without SClib. The parameters of the potential are the same as in Fig. 1. All the scripts were tested in the same machine, a notebook with a 2.4 Ghz core i5 processor (dual core) and 8 GB of RAM.}\end{table}

In \cite{Bra14} SChroe.py has been used to evaluate the relativistic corrections to
the mass spectrum of quarkonium in the long-distance regime. In that paper the
relativistic corrections $V^{(1/m)}$ and $V^{(1/m^2)}$ appearing in
(\DUrole{ref}{mass}) were evaluated assuming the hypothesis that in the long-distance
regime the interaction between the quark and the antiquark can be described by
a string. In Fig. 2 we show some of the energy levels (masses) corresponding to
the string spectrum. It is noteworthy to mention that all the numerical
calculations and plots of that paper were done with IPython using the SciPy
library.\begin{figure}[htb]\noindent\makebox[\columnwidth][c]{\includegraphics[width=\columnwidth]{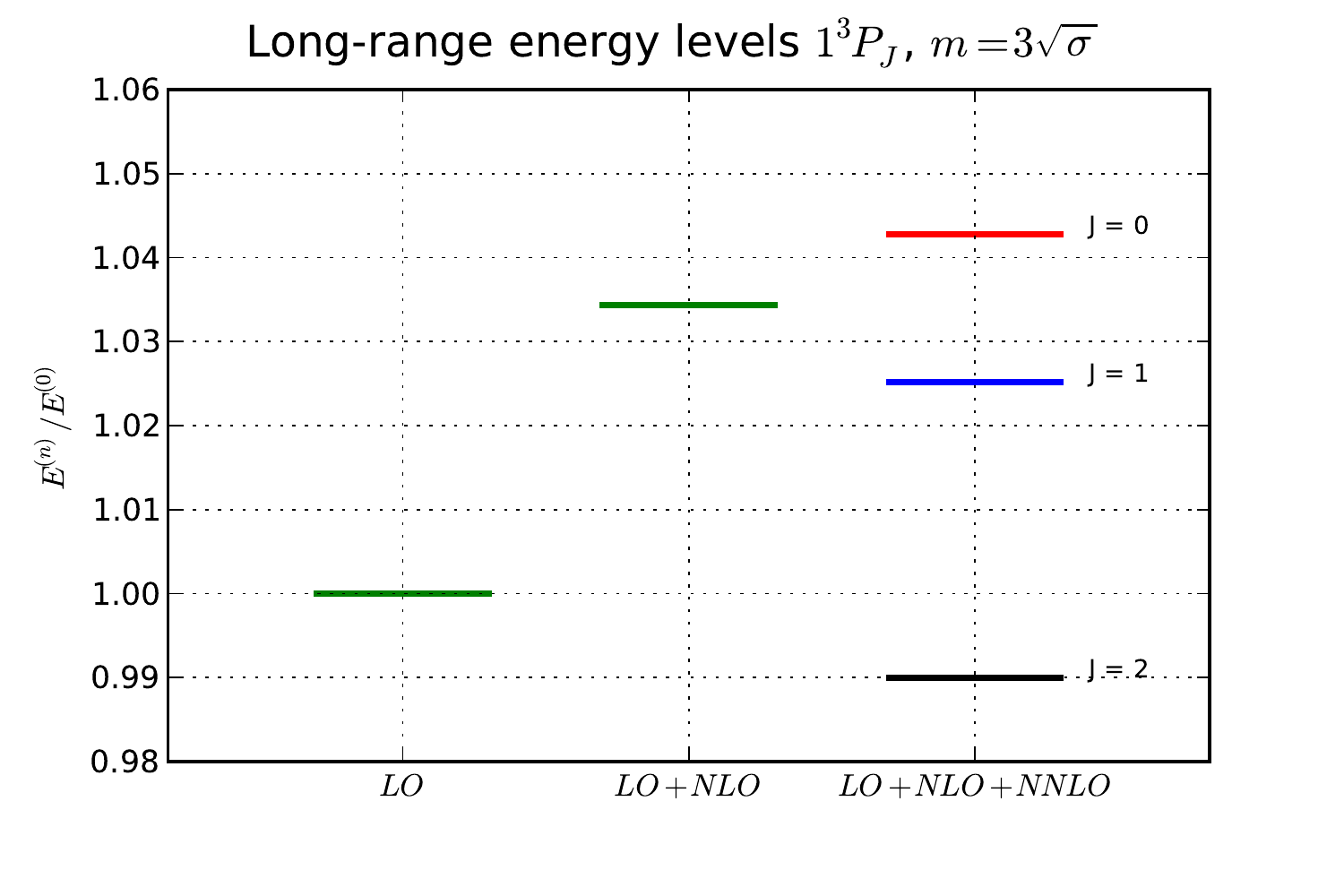}}
\caption{Long-range energy levels of the first triplet quarkonium state. The lines
are calculated from eq. (\DUrole{ref}{mass}) using the relativistic corrections
derived from the string hypothesis \cite{Bra14a}. The leading order (LO)
correspond to eq. (\DUrole{ref}{cornell}) setting $a=0$ and $k=1$ (in
the plot labeled $\sigma$) and $m=3\sqrt{k}$.  This plot shows
the relative size of the next-to-leading-order (NLO) correction (the term
proportional to $1/m$ in the r.h.s. of eq.  (\DUrole{ref}{mass})) and the
newly computed next-to-next-to-leading-order (NNLO) corrections (the terms
proportional to $1/m^2$). For more details see \cite{Bra14a}.}
\end{figure}

An application in which the speed of SChroe.py plays an important role is
fixing the parameters of the potential given some experimental input. For
instance, consider the problem of finding the parameters $a$ and
$k$ of eq.  (\DUrole{ref}{cornell}) together with $m$, given the
experimental values of the masses of three different quarkonium states. If
relativistic corrections are included, in order to find the parameters  we must
solve a system of three equations like eq. (\DUrole{ref}{mass}). For each probe value
of $(a,k,m)$ we have to find the eigenvalues and reduced wavefunctions of
eq. (\DUrole{ref}{reduced}) and then with these values evaluate the sums and integrals
in (\DUrole{ref}{mass}). A parameter fixing of this type was necessary to implement in
\cite{Bra14b}. The implementation has been carried out using SChroe.py together
with a mixture of C and SciPy functions using SClib to link both environments\DUfootnotemark{id43}{id44}{7}.%
\DUfootnotetext{id44}{id43}{7}{
Some of the code will be available once the paper appears online.}

Another related computational problem that arises from the study of heavy
quarkonium hybrids, bound-states composed by a quark-antiquark pair plus an
exited gluon, is to solve a system of $N$ Schrödinger-like coupled
equations.  Explicitly the system to solve reads\begin{equation}
\label{coupled}
\left(-\frac{\delta_{ij}}{m}\frac{d^2}{dr^2}+V_{ij}(r,l)\right)u_{j,(n,l)}(r)=E_{n,l}\,u_{i,(n,l)}(r),
\end{equation}where $i = 1,2,..N$ and the angular momentum dependence has been included
in the potential matrix. A method to solve this equation for the case
$N=2$ has been implemented in \cite{Ber14}. The method relies on an extension
of the nodal theorem \cite{Ama95} and convergence conditions for the components of
the vector wavefunction $u_{j,(n,l)}(r)$. The extension of the nodal
theorem states that the number of nodes of the determinant of the matrix
$U_{n,l}(r)$, whose columns are $N$ lineal-independent solutions of
eq. (\DUrole{ref}{coupled}), is equal to $n$. The procedure then consist in a
scan of values $E_{n,l}$; in each step the set of equations
(\DUrole{ref}{coupled}) is solved and the nodes of $|U_{n,l}(r)|$ are counted for
a large enough interval of $r$. As in the one-dimensional case, if
$E_{n,l}$ approached to an eigenvalue the components of
$u_{j,(n,l)}$ converge for large $r$. In the solution presented in
\cite{Ber14} the performance-intensive parts of the implementation rely on C
functions linked to the IPython interface trough SClib.

As an example of the application of the method implemented in \cite{Ber14}, in Fig.
3 we show the results for the search of the first two eigenvalues and
wavefunctions with the matrix potential given by\begin{equation}
\label{matrixpotential}
V_{ij}(r,l) = \begin{pmatrix} \frac{l(l+1)+2}{mr^2}+F_0(r) & -\frac{2\sqrt{l(l+1)}}{mr^2} \\ -\frac{2\sqrt{l(l+1)}}{mr^2} &  \frac{l(l+1)}{mr^2}+F_1(r) \end{pmatrix}
\end{equation}where\begin{equation*}
F_i(r)=\ln(a_i+b_ir).
\end{equation*}\begin{figure}[htb]\noindent\makebox[\columnwidth][c]{\includegraphics[width=\columnwidth]{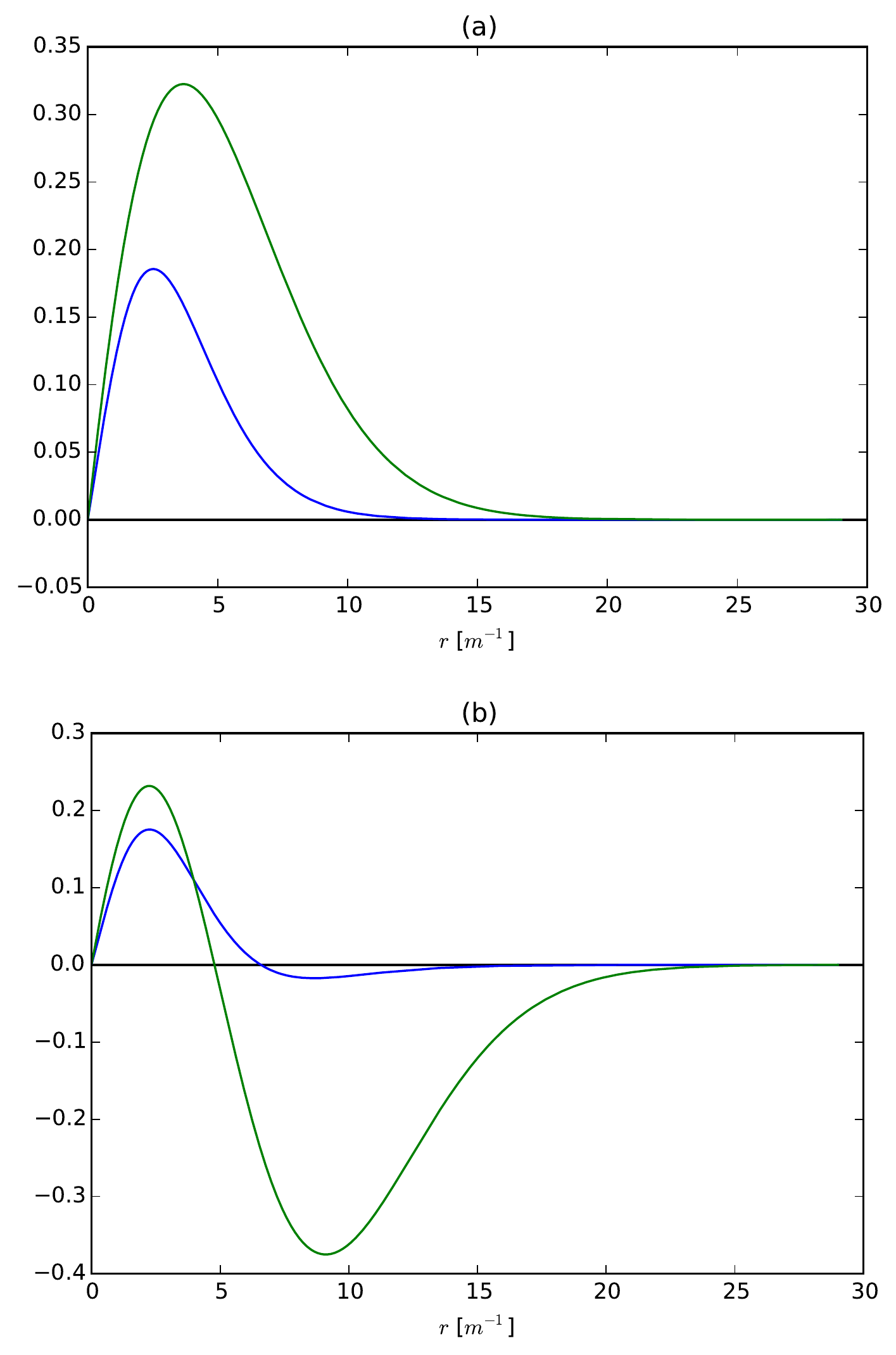}}
\caption{Solutions for the components of the vector wavefunction $u_{n,l}(r)$
for the first two eigenvalues ($l=1$) of eq. (\DUrole{ref}{coupled}) with the
matrix potential given in (\DUrole{ref}{matrixpotential}). We have used
$m=1$, $a_0=1$, $b_0=0.5$, $a_1=2$ and
$b_1=0.1$. The eigenvalues are $E_{n=0,l=1} = 1.01727\,m$ for
Fig. (a) and  $E_{n=1,l=1} = 1.18789\,m$ for Fig. (b).}
\end{figure}In all the applications described in this section the combination of SClib and
the SciPy library within an interactive environment like IPython provided a powerful framework based entirely on open source software for solving problems that require a high
performance and visualization tools.

\section{Application in Control Engineering%
  \label{application-in-control-engineering}%
}
\begin{figure}[]\noindent\makebox[\columnwidth][c]{\includegraphics[width=\columnwidth]{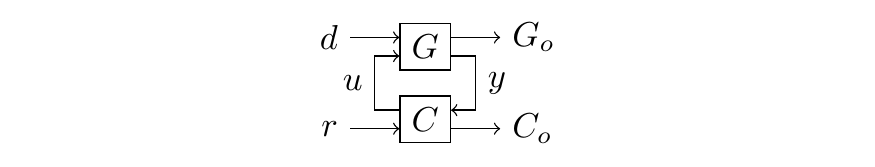}}
\caption{General scheme of a control system.}
\end{figure}

Most control systems have the structure depicted in Fig 4.  $G$ is the
plant, it represent the natural phenomena we wish to control.  We usually
describe it using ordinary differential equations:\begin{equation}
\label{eq:dxdt}
G:\;\left\{
    \begin{array}{rl}
        \frac{dx}{dt} &= f(x,u,d)\\
        y & = c(x,u,d).
    \end{array}
    \right.
\end{equation}$x$ represents the internal state of the plant and $y$ its output
(the measurements). $d$ is an independent variable, usually not
measurable, named the perturbation and $u$ is the actuation: the degree
of freedom used by the controller $C$ to achieve the control goal
$r$. In general the controller is a function of the measurements and the
reference $r$:\begin{equation*}
C:\;u = \pi(y,r),
\end{equation*}but it also may comprise internal states. They are commonly used to reconstruct
the state $x$ out of the history of $y$ and $u$. The latter
systems are called state observers and the whole is called feedback control.

We use SClib to put together a simulator for these kind of systems.  Both the
system derivatives $f(\cdot)$ and the control $\pi(\cdot)$ are
written in C and are evaluated using SClib. As stated before, the system state
represents a natural phenomena, therefore it is natural to describe it as a
continuous time variable, as eq. (\DUrole{ref}{eq:dxdt}) suggests. To calculate the
system state we have to solve this equation. In our simulator this is achieved
using numerical methods, namely the integration routines available in
\emph{scipy.integrate}. On the other hand, the controller is usually
implemented in a real-time computer, which can only sample $y$ at a fixed
interval (called $h$): it is a discrete-time system.  This means, that
the simulator only needs to evaluate $\pi(\cdot)$ at given times.

Traditional controllers took the form of linear filters, which could even be
implemented using analog circuitry. As control techniques and requirements
advance, more complex controllers are devised. Many modern control techniques
are based on optimization methods. Time-optimal controllers, for example,
require the solution of an usually very complex optimization problem, to find a
control $u$ that leads the system state $x$ towards its target $r$ in minimum
time \cite{Gru11}:\begin{equation}
\label{feedback}
u ^*= \pi^*(x)=\underset{\pi\in U,\, x\in X}{\operatorname{argmin}}\left\lbrace T_{x}(u)\right\rbrace.
\end{equation}Here $T_x(u)$ is the time required to lead $x$ towards its target
and $X$ and $U$ are the regions where we want $x$ and
$u$ to be confined, they constitute the constraints for the control
problem.  These kind of controllers require exhaustive computation and it is
natural to implement them in C.

For motivation, we present the results for a minimum-time control strategy for
a relatively simple and well known problem, the double integrator
\cite{Fu13}:

\begin{equation}
\label{eq:di}
\frac{d}{dt}\left[
    \begin{array}{c}
        x_0\\
        x_1
    \end{array}
\right]
=
\left(
\begin{array}{c}
    \nicefrac{u}{\tau_0}\\
    \nicefrac{x_0}{\tau_1}
\end{array}
\right).
\end{equation}The relevance of this system lays in that it models many mechanical systems:
$u$, $x_0$ and $x_1$ may represent acceleration, speed and
position, for example.

Fig. 5 presents a minimum time control strategy for this system.\begin{figure}[htb]\noindent\makebox[\columnwidth][c]{\includegraphics[width=\columnwidth]{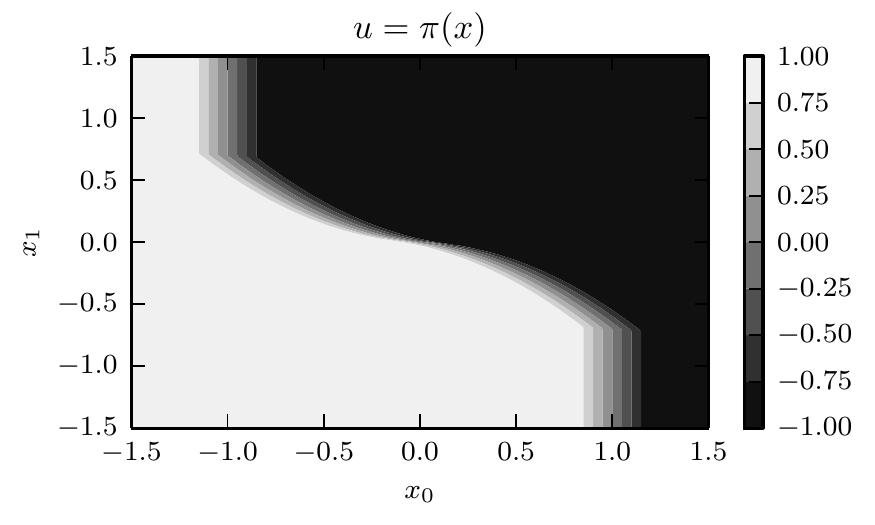}}
\caption{Time optimal control for the double integrator considering
$\tau_0=\tau_1=5$, $u\in[-1,1]$, $h=1$ and $x\in[-1,
1]\times\mathbb{R}$.}
\end{figure}

The form of $\pi(x)$ for this case reveals its non-linear nature.

Fig. 6 presents the trajectory developed by the state using this control
strategy and random initial conditions.\begin{figure}[htb]\noindent\makebox[\columnwidth][c]{\includegraphics[width=\columnwidth]{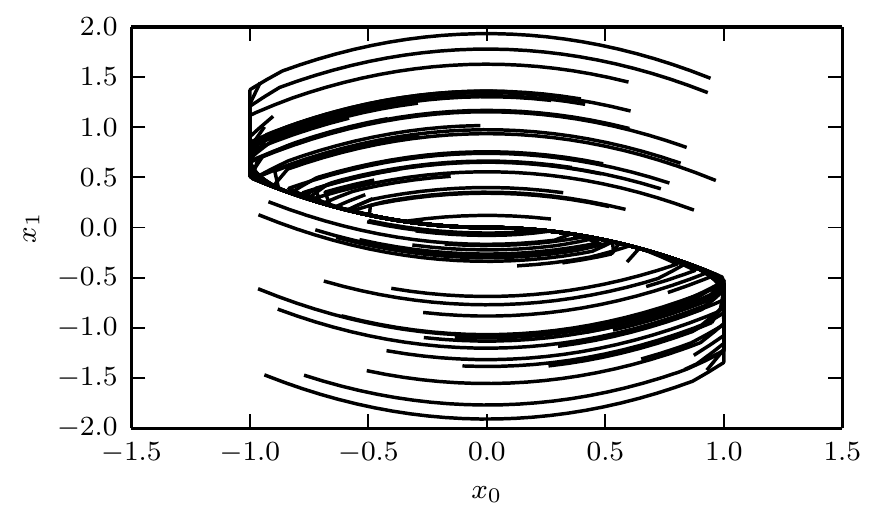}}
\caption{Time optimal trajectories for the double integrator, with random initial
conditions.}
\end{figure}

These results were obtained using SClib and the devised simulator. The example
code provided with SClib is ready to reproduce them.

The main advantage we obtained from this work was that, since we were using a
Linux based real time system in our test rig, we could use exactly the same
code for the simulations and the experimental tests.  Another feature of this
work is that it effectively replaces Simulink in all of our use cases using
only free software.

\section{Final Remark%
  \label{final-remark}%
}

We hope the applications of SClib scope beyond the ones listed in this paper
since we believe it provides a simple but powerful way to boost Python
performance.

\section{Acknowledgments%
  \label{acknowledgments}%
}

H.M. acknowledges financial support from DAAD and the TUM Graduate School
during the realization of this work.

\end{document}